\documentclass[aip,jcp,
 amsmath,amssymb,
 reprint,%
floatfix]{revtex4-2}

 \usepackage[utf8]{inputenc}
 \usepackage{multirow}
 \usepackage{soul}

\usepackage{chemformula} 
\usepackage[T1]{fontenc}
\makeatother

\newcommand{\iceIH}{ice I$_{\text{h}}$}


\begin{document}

\title{Hexagonal ice density dependence on interatomic distance changes due to nuclear quantum effects}

\author{Lucas T. S. de Miranda}
\affiliation{Institute of Theoretical Physics, São Paulo State University (UNESP), Campus São Paulo, Brazil}
\author{Márcio S. Gomes-Filho}
\affiliation{Centro de Ciências Naturais e Humanas, Universidade Federal do ABC, 09210-580, Santo André, São Paulo, Brazil}
\author{Mariana Rossi}
\affiliation{Max Planck Institute for the Structure and Dynamics of Matter, Luruper Chaussee 149, 22761 Hamburg, Germany}
\author{Luana S. Pedroza}
\affiliation{Instituto de F\'isica, Universidade de S\~ao Paulo, SP 05508-090, Brazil}
\author{Alexandre R. Rocha}
\affiliation{Institute of Theoretical Physics, São Paulo State University (UNESP), Campus São Paulo, Brazil}
\affiliation{Max Planck Institute for the Structure and Dynamics of Matter, Luruper Chaussee 149, 22761 Hamburg, Germany}
\email{alexandre.reily@unesp.br}








\begin{abstract}

Hexagonal ice ($\rm{I_h}$), the most common structure of ice, displays a variety of fascinating properties. Despite major efforts, a theoretical description of all its properties is still lacking. In particular, correctly accounting for its density and interatomic interactions is of utmost importance as a stepping stone for a deeper understanding of other properties. Deep potentials are a recent alternative to investigate the properties of {\iceIH}, which aims to match the accuracy of \textit{ab initio} simulations with the simplicity and scalability of classical molecular dynamics. This becomes particularly significant if one wishes to address nuclear quantum effects. In this work, we use machine learning potentials obtained for different exchange and correlation functionals to simulate the structural and vibrational properties of {\iceIH}. We show that most functionals overestimate the density of ice compared to experimental results. Furthermore, a quantum treatment of the nuclei leads to even further distancing from experiments. We understand this by highlighting how different inter-atomic interactions play a role in obtaining the equilibrium density. In particular, different from water clusters and bulk water, nuclear quantum effects lead to stronger H-bonds in {\iceIH}.


\end{abstract}

\maketitle

\section{Introduction}

Hexagonal ice (\iceIH), commonly known as ordinary ice, exhibits a range of remarkable properties that are essential to life and play a significant role in regulating Earth's climate.\cite{Petrenko99:book} 
Ice I$_{\text{h}}$ also plays an important role in several areas such as atmospheric and environmental chemistry, cryobiology, materials science, and engineering.\cite{moberg2017molecular} 
It has a tetragonal structure originally proposed by Pauling,\cite{pauling1935structure} based on experiments by Dennison,\cite{dennison1921crystal} Bragg\cite{bragg1921crystal} and Barnes.\cite{barnes1929crystal} 
Despite its simple chemical structure, {\iceIH} displays an anomalous density behavior\cite{Chen17} and some properties are still not fully understood, particularly those relating to the hydrogen bond network that drives cohesion in both solid and liquid phases of water.\cite{santos2020density}

Together with water, ice has been the subject of extensive theoretical studies from a microscopic perspective, with methods such as Monte Carlo and molecular dynamics (MD) simulations using empirical models for water-water interactions.\cite{jorgensen1983comparison, berendsen1987missing, mark2001structure, abascal2005potential} 
Although some classical models can yield good results for ice phases,\cite{jorgensen1983comparison, berendsen1987missing} like TIP4P/Ice,\cite{abascal2005potential} their accuracy and transferability are often limited, posing challenges for accurately reproducing experimental data across diverse conditions with a single parameter set.

In that sense, \textit{ab initio} molecular dynamics (AIMD) simulations based on density functional theory (DFT)\cite{hohenberg1964inhomogeneous, kohn1965self} has become the workhorse of simulations of solids and liquids as it provides a balance between accuracy and possible system sizes. \cite{morrison1997ab, english2015structural}
While AIMD can be highly accurate, it is typically restricted to smaller systems (hundreds of atoms) and short simulation times (tens of picoseconds), with outcomes that depend on the quality of the exchange-correlation (XC) functional employed. This is particularly a troublesome in the case of water molecules,   where, due to a subtle balance between hydrogen bonds  and van der Waals (vdW) interactions, most XC functionals do not properly describe all the features of liquid water.\cite{gillan2016perspective} Recently, Montero \textit{et. al.}\cite{montero2024density} investigated a variety of XC functionals for water and {\iceIH} systems and showed there is no functional that agrees well with  experimental results for the equilibrium density for both systems.

Furthermore, for water systems, nuclear quantum effects (NQEs) have been shown to significantly alter structure and dynamics, affecting hydrogen bond strength and structural properties.\cite{Morrone2008:PRL,MichaelidesPIMD, NQEreview, pamuk2012anomalous, ganeshan2013simulation} 
NQEs has also been shown to improve the structural properties of different ice phases.\cite{pamuk2012anomalous, hernandez2005quantum, herrero2011isotope}

These effects are typically included in simulations by using path integral molecular dynamics (PIMD)\cite{chandler1981exploiting}. Cheng \textit{et. al.} explored the thermodynamic characteristics of liquid water by employing DFT at the hybrid-functional level (revPBE0-D3), taking into account NQEs. Their findings indicated that these effects resulted in approximately an 1\% increase in the density of liquid water, hexagonal ice, and cubic ice compared to simulations that treated nuclei classically.\cite{Cheng2019:PNAS} 
Similar results were also observed for liquid water calculated using the metaGGA XC functionals SCAN~\cite{xu2020isotope} and SCAN0~\cite{zhang2021modeling:scan0}. 
However, combining AIMD and PIMD is computationally demanding and typically limited to short timescales due to the complexity of incorporating NQEs.

Machine learning potentials, particularly deep neural network force fields, also known as deep potentials (DP)~\cite{wang2018deepmd, Zhang18, Wen22,DeepMDreview2023:JCP}, offer an efficient way to represent the potential energy surface, and thus are appealing for performing molecular dynamics (MD) by integrating the accuracy of \textit{ab initio} calculations with computational efficiency comparable to classical MD.\cite{zhang2021phase, wang2022machine} Recently, DP were used to investigate some properties of {\iceIH},\cite{piaggi2021phase} and super ionic ice.\cite{matusalem2022plastic} With the SCAN quality, the melting temperature of \iceIH is around 310 K.\cite{piaggi2021phase} 
This is advantageous, particularly when considering simulations that include NQEs in water systems,\cite{Torres21, ko2019isotope, zhang2021modeling:scan0, zhang2020isotope, xu2020isotope} enabling longer and larger-scale calculations - both in terms of number of atoms and total simulation time - compared to traditional AIMD with PIMD.

This work aims to elucidate the role of including NQEs on the structural properties of {\iceIH}. This is achieved through a detailed analysis of the equilibrium density, the inter- and intra-atomic distances, and the vibrational frequencies within water molecules. To pursue this goal, we conducted simulations employing Deep Potential Molecular Dynamics (DP-MD) and Deep Potential Path Integral Molecular Dynamics (DP-PIMD) across a range of DP models trained on different XC potentials. We observe that the DP-MD results can be compared to AIMD calculations. The inclusion of NQEs leads a slight increase of the median of the OH covalent bond length, causing a red shift in stretching frequencies. As a consequence, proton sharing is observed, akin to results for bulk water.\cite{ceriotti2013nuclear,ceriotti2016nuclear} At the same time, this is in direct competition with decreasing intermolecular distances due to strengthening of the hydrogen bonds. 
This results in an overall increase of equilibrium density for all XC functionals, further away from experiments. This indicates that all functionals, either GGA-type or SCAN tend to overbind water molecules in ice, and NQEs further increase the strength of the hydrogen bonds.

\section{Methods}

\begin{figure}[!ht]
    \centering
    \includegraphics{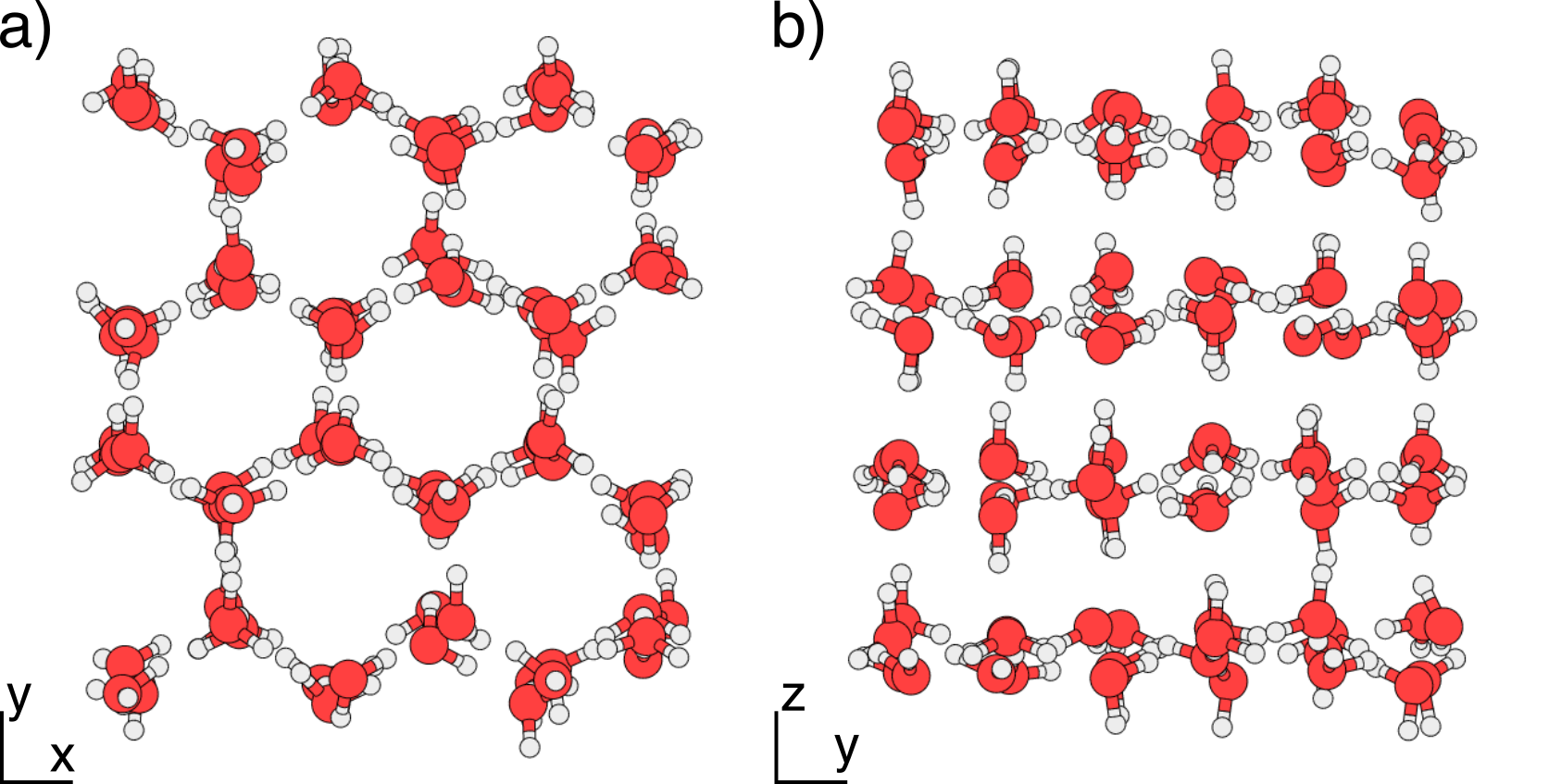}
    \caption{Top (a) and side (b) view of a rectangular cell for \iceIH~structure, where the red and gray spheres represent the oxygen and hydrogen atoms, respectively.}
    \label{fig:ice}
\end{figure}

Our investigation into the role of NQEs in {\iceIH} followed a structured workflow consisting of five key stages. In the first stage, we trained a preliminary deep potential (DP) models for several exchange-correlation (XC) functionals—PBE,\cite{perdew1996generalized} vdW-cx,\cite{BH} optB88-vdW,\cite{optB88} and SCAN\cite{SCAN-original}—using a DFT dataset from Torres \textit{et al.}\cite{Torres21, GomesFilho23} 
PBE is one of the most widely used XC functionals, despite its known tendency to overstructure bulk water.\cite{lin2012structure}
vdW-cx and optB88-vdW functionals incorporate van der Waals (vdW) interactions, essential for refining water’s structural and dynamical properties.\cite{lin2009importance, morawietz2016van} SCAN, a member of the metaGGA functional family, was included due to its recent success in accurately modeling bulk water.\cite{chen2017ab}

The second stage involved performing isobaric-isothermal (NPT) ensemble DP-MD simulations with the preliminary models to generate ice configurations, which were then added to the training data set.
In the third stage, DFT energies and forces were calculated for each selected ice configuration using each of the XC functionals.
In the fourth stage, we retrained and validated each DP model by incorporating the newly generated ice configurations into the training set.

Finally, in the fifth stage, the refined DP models were used to perform NPT DP-MD and DP-PIMD simulations. 
All simulations were performed using a 96 molecule {\iceIH} box with randomized hydrogen bond orientations to ensure zero net electric dipole moment within the simulation cell (see Fig. \ref{fig:ice}). Each step is described in further detail below, and more information can be found in the Supplementary Material (SM), including a schematic figure of the workflow. 

\subsection{Training deep potentials}

The preliminary models were developed from a dataset encompassing a broad range of liquid water geometries. These configurations were extracted from uncorrelated snapshots of long classical MD simulations that sampled various system conditions, including diverse densities and temperatures. For each XC functional, the initial dataset consisted of DFT-calculated energies and forces for 18,000 snapshots, each composed of 64-water-molecule box. Detailed information about the initial dataset can be found in Torres \textit{et. al.}.\cite{Torres21}

We then performed NPT DP-MD simulations for each preliminary DP model on {\iceIH} composed of 96 water molecules, at T = 220 K and P = 0.1 MPa, with a 2 ns production run using a 0.25 fs timestep in the LAMMPS code.\cite{Lammps22} The chained Nosé-Hover\cite{tuckerman1993efficient} and barostat with three chains were used for these simulations. From each simulation, 1,000 configurations were extracted at uniform time intervals (2 ps) and added to the initial dataset. This improvement of the DP model can be interpreted as a transfer learning approach.\cite{TransferDEEPMD}

Single-point DFT calculations were then performed on each ice configuration using the SIESTA\cite{artacho2018siesta} code for the PBE, vdW-cx, and optB88-vdW functionals. These calculations employed norm-conserving pseudopotentials\cite{troullier1991efficient}, a 500 Ry mesh cutoff, and an optimized double-$\zeta$ basis set with polarization orbitals for valence electrons.\cite{corsetti2013optimal} 
SCAN single-point calculations were computed using VASP\cite{kresse1996efficient}, with a 118 Ry energy cutoff and the projected augmented wave (PAW) method for core-valence interactions.\cite{kresse1999ultrasoft}

We used the current GPU version of the software package DeePMD-kit v2.1.3.\cite{wang2018deepmd, DeepMDreview2023:JCP, lu202186} to generate deep neural network potentials. The Deep Potential-Smooth Edition descriptor was used in which all relative coordinates (distances and angles) are used to build the descriptor\cite{zhang2018end}, within a cutoff radius of 6.0 {\AA} and a smoothing ratio of 0.5 {\AA}.  We first selected 90\% of the configurations as the training set and remaining 10\% as the test set for assessment.

The loss function, comprising mean squared errors in energies and forces, was minimized using the Adam stochastic gradient descent method over $7\times10^6$ training steps. This training procedure was repeated for each XC functional, producing a distinct DP model for each.

To evaluate the quality of the DP models, we performed two validations: (i) \textit{a priori} validation, where errors in energy and force predictions from DP models were compared to the DFT reference values using the test dataset, and (ii) \textit{a posteriori} validation, where we examine if the coordinates generated from DP-MD simulations produce structures with reasonable errors for energies and forces. For each XC functional, we obtain small errors on the testing dataset, below $45$ meV/\AA~(forces) and $0.5$ meV/atom (energy), well within the acceptable range for a reliable DP model.\cite{Wen22} 
It is essential to note that the \textit{a priori} validation involved testing the models on both liquid and solid configurations. At the same time, during the \textit{a posteriori} validation, the DP models are tested on 200 {\iceIH} configurations with different densities, collected during the production run, yielding even lower errors. This suggests that the fitted potentials are accurate enough. Given the lower error values we observed in the \textit{a posteriori} validation, it was not necessary to include more training data points in the training data set. Parity plots and the errors for each XC functional are presented in the SM.

\subsection{Production run: Molecular Dynamics}

After training a deep neural network for each XC functional, we conducted two independent NPT molecular dynamics simulations: one with nuclear quantum effects (NQEs) included (DP-PIMD) and one without it (DP-MD). 
These simulations were performed on a proton-disordered \iceIH~structure composed of 96 water molecules, as shown in Figure \ref{fig:ice}. 
For all simulations, systems were initially equilibrated over 100 ps, followed by 500 ps of production runs with a 0.25 fs timestep, and data was collected every 0.05 ps. 
The simulations were conducted at a specific point on the water phase diagram (T = 220 K and P = 0.1 MPa), allowing direct comparisons with experimental results.\cite{soper2000radial}

All simulations used the i-PI code.\cite{kapil2019pi} 
We used the Generalized Langevin Equation (GLE) thermostat\cite{ceriotti2010colored} for DP-MD (optimal sampling), and for DP-PIMD simulations we used the PIGLET\cite{ceriotti2012efficient, ceriotti2009nuclear} thermostat with 8 beads and ratio of $\rm{\omega_{max}/\omega_{min}=10^4}$ to include the NQEs.  
Additional details on the convergence with respect to the number of beads, further testing, and the procedure for obtaining vibrational spectra are provided in the SM.

\section{Results and Discussion}

The equilibrium densities of {\iceIH} for each XC functional are listed in Table~\ref{tab: density}. 
We observe that for the optB88-vdw, vdW-cx and SCAN functionals, the DP-MD equilibrium density shows a significant deviation from experimental values, exceeding them by more than 5\%.
However, for PBE, the equilibrium density is closer to the experimental data. As we will discuss later this is most likely a spurious effect. 
PBE is known to ovestructure water as it tends to overbind, and it is probably benefiting from this property.

\begin{table}[!ht]
    \centering
        \begin{tabular*}{\columnwidth}{@{\extracolsep{\fill}}ccc@{}}\hline\hline
& \multicolumn{2}{c}{Density ($\rm{g/cm^3}$)} \\\hline\hline
 XC & DP-MD & DP-PIMD \\ \hline\hline
 PBE & 0.937 (2.18\%)& 0.961 (4.80\%) \\
 vdW-cx & 0.981 (6.98\%) & 1.013 (10.47\%) \\
 SCAN & 0.974 (6.22\%)& 1.001 (9.16\%)\\
 optB88 & 0.964 (5.51\%)& 0.984 (7.31\%) \\\hline\hline
 Exp.\cite{soper2000radial} & \multicolumn{2}{c}{0.917}\\ \hline\hline
    \end{tabular*}
    \caption{Equilibrium density for each XC functional for both methods. The values in parenthesis are the deviations from experimental data.\cite{soper2000radial} The standard deviation of each equilibrium density presented is of order $\pm 0.009$ $\rm{g/cm^3}$. The graphs of density and also the related errors are presented in SM.}
    \label{tab: density}
\end{table}

Notably, the equilibrium densities obtained here are similar to those obtained from AIMD for PBE (0.936 $\rm{g/cm^3}$) and SCAN (0.964 $\rm{g/cm^3}$) as seen in Ref. \cite{Chen17}. When compared to previous works\cite{Torres21}, we find that PBE predicts ice to be denser than water, failing to reproduce the well-known density anomaly of water.\cite{Chen17} This anomalous behavior, where ice is less dense than water, is correctly captured by the other XC functionals.\cite{Torres21}

The inclusion of NQEs increases the equilibrium density for all XC functionals tested. As mentioned before, this density increase (volume contraction) was also observed in different DFT-based results for shorter simulation times.\cite{Cheng2019:PNAS, xu2020isotope, zhang2021modeling:scan0, pamuk2012anomalous} The density increase observed in our calculations ranges from about 1\% to 3\%, depending on the model, as shown in Table \ref{tab: density}.

\begin{figure*}[!ht]
    \centering
    \includegraphics[width=\textwidth]{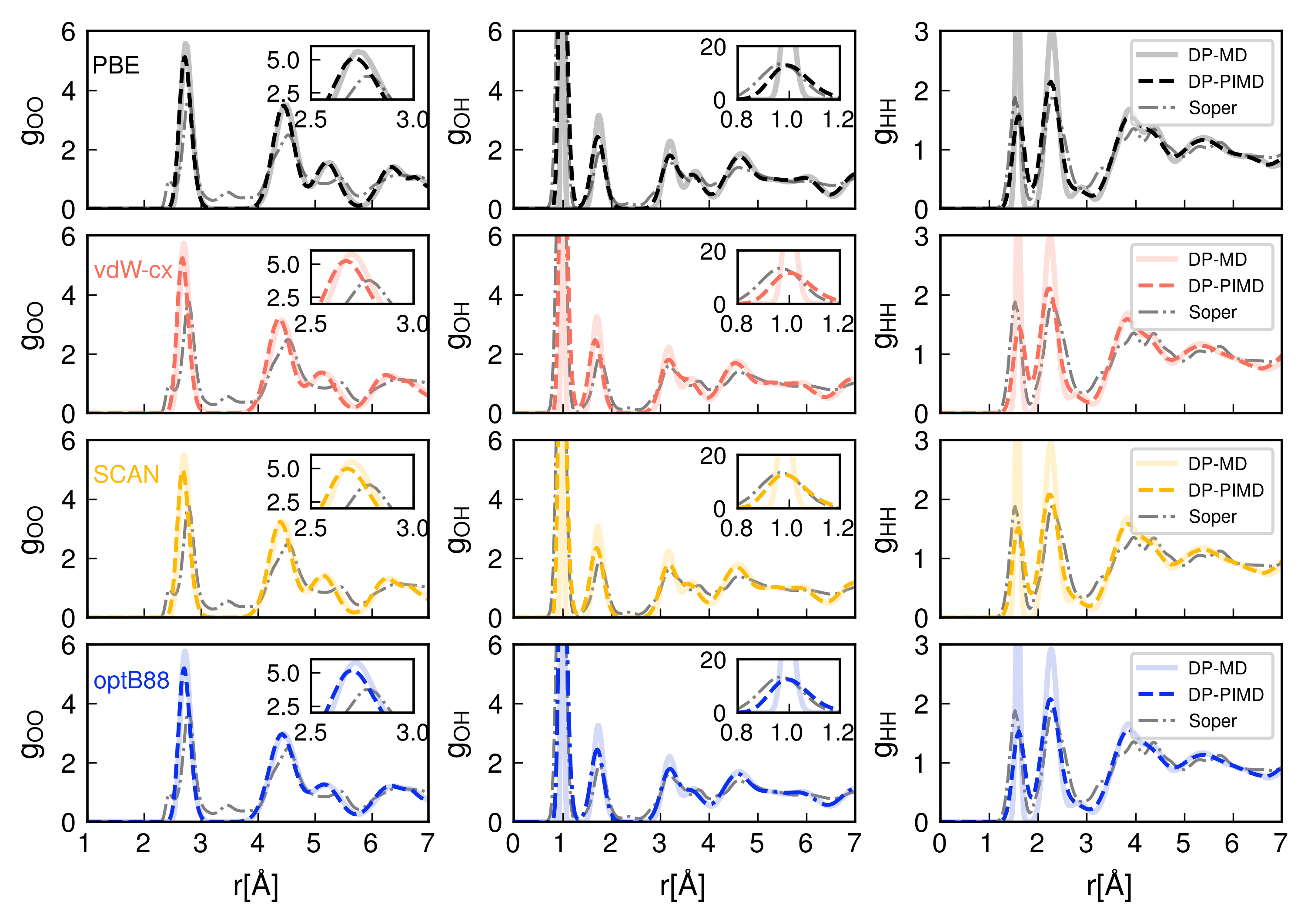}
    \caption{RDFs for oxygen-oxygen (g$\rm_{OO}$) (left panel), oxygen-hydrogen (g$\rm_{OH}$) (center panel), and hydrogen-hydrogen (g$\rm_{HH}$) (right panel) obtained via DP-MD (dashed lines) and DP-PIMD (solid lines) simulations and the experimental result from  Soper\cite{soper2000radial} (dash-doted line), for different XC functionals (up to bottom: PBE, vdW-cx, SCAN and optB88-vdW). The inset graphs show the first peak. }
    \label{fig: rdf}
\end{figure*}

\begin{table}[!ht]
    \centering
        \begin{tabular*}{\columnwidth}{@{\extracolsep{\fill}}cccccc@{}}\hline\hline
   XC & Method & $\rm{\epsilon_{OO}}$ (\%) & $\rm{\epsilon_{OH}}$ (\%) & $\rm{\epsilon_{HH}}$ (\%) & $\rm{\epsilon_{avg}}$ (\%)\\ \hline\hline
        \multirow{2}{*}{PBE} & DP-MD & 81.6 & 77.7 & 86.9 & 82.1\\
        & DP-PIMD & 81.1 & 87.5 & 89.9 & 86.2 \\\hline\hline
        \multirow{2}{*}{vdW-cx} & DP-MD & 81.1 & 76.5 & 86.2 & 81.3\\
        & DP-PIMD & 78.3 & 84.5 & 87.6 & 83.5\\\hline\hline
        \multirow{2}{*}{SCAN} & DP-MD & 80.9 & 78.2 & 86.5 & 81.9 \\
        & DP-PIMD & 79.3 & 87.9 & 88.5 & 85.2 \\\hline\hline
        \multirow{2}{*}{optB88} & DP-MD & 83.7 & 78.4 & 87.3 & 83.1\\
        & DP-PIMD & 82.5 & 88.0 & 89.6 & 86.7\\\hline\hline
    \end{tabular*}
    \caption{The performance evaluation of each XC functional is based on a comparison between their respective RDFs results and the experimental data obtained from equation~(\ref{eq:errorRDF}) in percent.}
    \label{tab:errorRDF}
\end{table}

To investigate more closely the increase in equilibrium density for {\iceIH}, we first calculated the radial distribution function (RDFs) for oxygen-oxygen ($g_{OO}$), oxygen-hydrogen ($g_{OH}$), and hydrogen-hydrogen ($g_{HH}$) pairs using each XC functional, as shown in Figure~\ref{fig: rdf}. Overall, the oxygen-oxygen RDFs show a shorter O-O distance in comparison with the experimental data. Moreover, the inclusion of NQE does not result in significant changes to the O-O curves; the exception is only on the position of the first peak, as shown in the inset of the first column. Furthermore, the experimental data for O-H and H-H RDFs deviates significantly from those obtained with DP-MD simulations, which show more structuring than the experimental data. 

The inclusion of NQEs, however, substantially improves structural predictions of these $g(r)$. Overall, we note smoothing of the curves, and the intensity of the maxima and minima in the RDFs are in good agreement with the experimental results.\cite{soper2000radial} In addition, it is interesting to note that this improvement of structural properties appears to make the curves less sensitive to the choice of XC functional (see Figure S8 in the SM).

\begin{figure*}[!ht]
    \centering
    \includegraphics{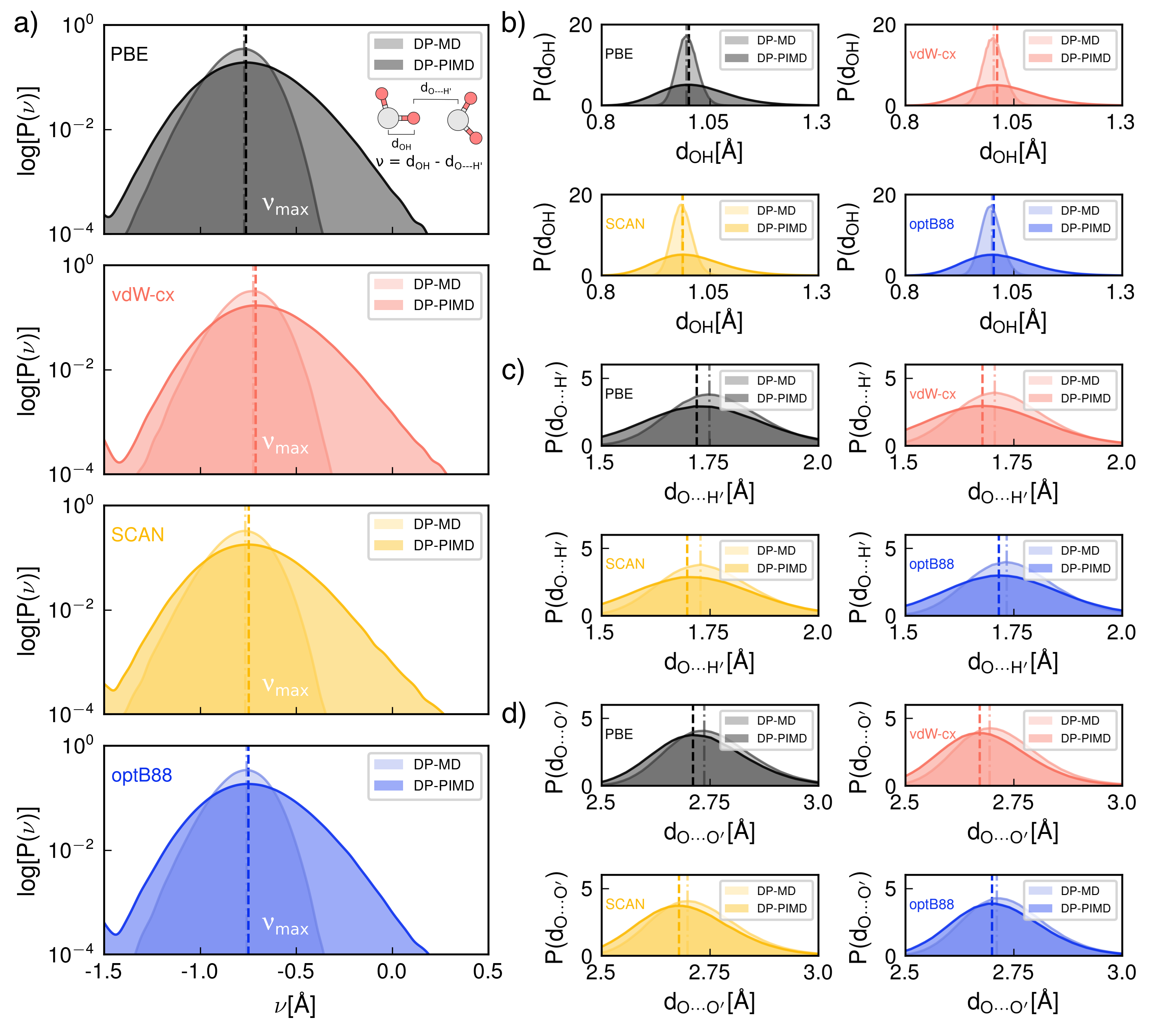}
    \caption{Distributions of a) proton transfer coordinate $\nu$ and bond distances of b) intramolecular oxygen-hydrogen ($\rm{d_{OH}}$), c) intermolecular oxygen-hydrogen ($\rm{d_{O\cdots H'}}$) and d) oxygen-oxygen ($\rm{d_{O\cdots O'}}$) obtained via DP-MD and DP-PIMD simulations for different XC functionals. For all distributions, the dashed (dash-doted) line represents the most probable value for the bond length of DP-PIMD (DP-MD) simulations.}  
    \label{fig: dist}
\end{figure*}

To quantify how closely each XC functional matches experimental RDFs, we used the quality metric proposed by Schran \textit{et al}.,\cite{Schran21}
\begin{equation} \label{eq:errorRDF}
    \epsilon_{ij} = 1 - \frac{\int_{0}^{+\infty} | g_{ij}^{\text{EXP}}(r) - g_{ij}^{\text{DP}} (r) |dr}{ \int_{0}^{+\infty} g_{ij}^{\text{EXP}}(r)dr +   \int_{0}^{+\infty} g_{ij}^{\text{DP}}(r)dr}
\end{equation}
where the sub-index $ij$ represents atomic pairs (O-O, O-H, and H-H). This metric provides a similarity measure between two RDFs ranging from 0 (most dissimilar) to 1 (identical).\cite{Schran21}

The values of $\rm{\epsilon_{ij}}$ for each XC functional are presented in Table~\ref{tab:errorRDF}, along with the averaged accuracy metric,  $\rm{\epsilon_{avg}= (\epsilon_{OO}+\epsilon_{OH}+\epsilon_{HH})/3}$. 
Notably, NQEs mainly affect the O-H interaction as seen in Table \ref{tab:errorRDF}, with an increase in accuracy by approximately 10\%. 
This improvement correlates with a softening effect on the first peak of the O-H RDF, indicating a closer alignment with experimental values. Similarly, NQEs lead to an improved fit in the H-H RDF.

It is worth noting that in the region around the third peak of $g_{OH}$, we observed a significant improvement by including NQEs when compared to the experimental data. This indicates that quantum fluctuations increase the configurational entropy of the ice structures, broadening the distributions, and bringing them closer to the observations. This phenomenon has also been recently reported for liquid water.\cite{xu2020isotope} 
Overall, incorporating NQEs generally increased RDF accuracy across all XC functionals. The only exception is the O-O RDF, where accuracy slightly decreased as NQEs shift the positions of the first two peaks to lower distances, reflecting the change in equilibrium density.

\begin{table*}[t]
    \centering
    \footnotesize
    \begin{tabular}{ccccccccc}\hline\hline
 & \multicolumn{2}{c}{$\nu_{max}$ (\AA)} & \multicolumn{2}{c}{$d_{OH}$ (\AA)} & \multicolumn{2}{c}{$d_{O\cdots H'}$ (\AA)} & \multicolumn{2}{c}{$d_{O\cdots O'}$ (\AA)}\\ \hline\hline
 XC  &  DP-MD & DP-PIMD & DP-MD & DP-PIMD & DP-MD & DP-PIMD & DP-MD & DP-PIMD  \\ \hline\hline
 PBE &  -0.772 & -0.762 & 0.995 & 1.001 & 1.748 & 1.719 & 2.736 & 2.710\\
 vdW-cx &  -0.725 & -0.712 & 1.003 & 1.011 &1.706 &1.677 &2.694 & 2.671\\
 SCAN &  -0.767 & -0.748 & 0.985 & 0.987 & 1.728 & 1.697 & 2.698 & 2.678\\
 optB88 &  -0.759 & -0.751 & 0.999 & 1.003 &1.733 & 1.715 &2.710 & 2.699\\\hline\hline
 Exp.  & & & \multicolumn{2}{c}{1.008\cite{Kuhs83}} & \multicolumn{2}{c}{1.750\cite{brill1967gitterparameter}} & \multicolumn{2}{c}{2.759\cite{Kuhs83}} \\ \hline\hline
\end{tabular}
    \caption{Spatial distributions of proton transfer coordinate $\nu$ maxima and inter atomic distances ($\rm{d_{OH}}$, $\rm{d_{O\cdots H'}}$ and $\rm{d_{O\cdots O'}}$) for DP-MD and DP-PIMD simulations.}\label{tab: nu}
\end{table*}

To understand the change in density associated with NQEs, we present in Figure \ref{fig: dist}a) the spatial distribution of the proton transfer coordinate $\rm{\nu = d_{OH} - d_{O\cdots H'}}$ - which serves as an indicator of a proton transfer event\cite{xu2020isotope}.
We also present distributions for intra- and intermolecular bond distances. Notably, NQEs broaden the peaks and show non-zero density at positive values of $\rm{\nu}$ similar to that which has been observed in liquid water \cite{ceriotti2013nuclear, xu2020isotope}.
Most importantly, the maximum of the distribution, $\rm{\nu_{max}}$, exhibits a slight decrease with longer distances shift as seen in the Figure \ref{fig: dist}a) and in Table \ref{tab: nu}. This observation is linked to the strengthening of intermolecular interactions since this peak is sensitive to the hydrogen bond network. This subtle effect likely results from the competition among various interatomic interactions.

Figures \ref{fig: dist}(b-d) show the distributions for the intramolecular O-H bond distance (${d_{OH}}$), the intermolecular O-H distance (${d_{O\cdots H'}}$), and the intermolecular O-O distance (${d_{O\cdots O'}}$), while Table \ref{tab: nu} lists the corresponding peak positions. 
We note that the average $\rm{d_{OH}}$ distances from both DP-PIMD and DP-MD simulations are comparable, though the DP-PIMD distribution is broader and slightly right-skewed, suggesting a weakening of the covalent bond. 
At the same time,  the $\rm{d_{O\cdots H'}}$ and $\rm{d_{O\cdots O'}}$ distance distributions are also broader but left-skewed in this case (see Figure \ref{fig: dist}c and Figure \ref{fig: dist}d respectively). This is in line with expectations, as NQEs elongate the covalent bonds and promote proton sharing in bulk water.\cite{pamuk2012anomalous, ceriotti2016nuclear}  
At the same time, Li {\it et al} have argued that the hydrogen bond in bulk water and water clusters is relatively weak, and thus tends to become weaker with the inclusion of NQEs \cite{doi:10.1073/pnas.1016653108}. For ice, however, we observe the opposite, {\it i.e.}, the average value for  ${d_{O\cdots H'}}$ and ${d_{O\cdots O'}}$ is significantly shorter, leading to the conclusion that the hydrogen bond is relatively strong in ice, and thus becomes stronger when NQEs are taken into account (see Figure S6). 

\begin{figure}[!ht]
    \centering
    \includegraphics[width=0.5\textwidth]{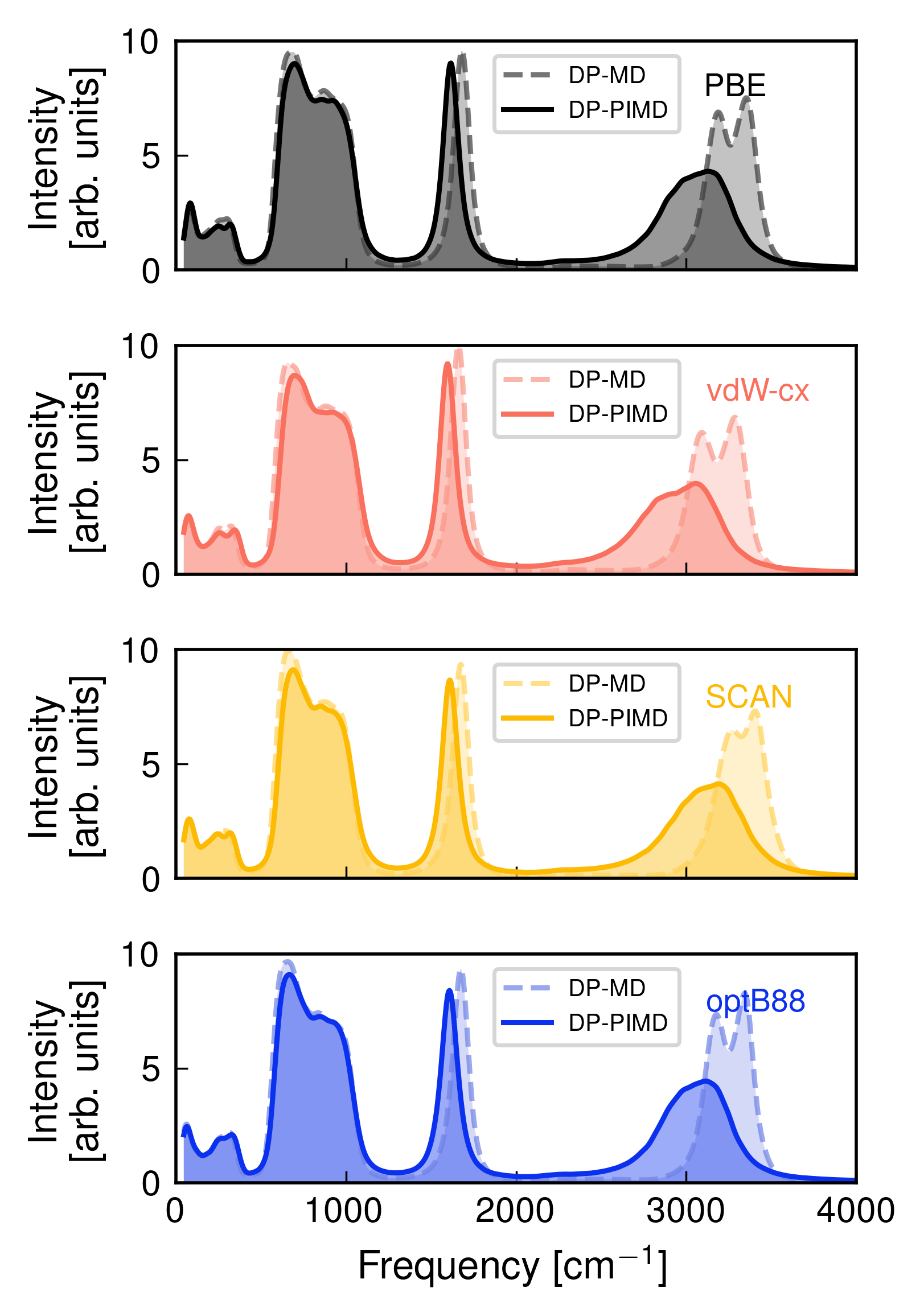}
    \caption{Vibrational spectra for \iceIH~ at 220 K obtained via path integral-based method, namely: thermostatted ring polymer molecular dynamics (TRPMD) simulations~\cite{Rossi14:JCP,Rossi14:JCPcommun} (solid lines): DP-PIMD. For comparison purposes, we also show the vibrational spectra obtained via classical MD simulations, where the number of particles N, the volume V, and energy E were kept constant during the simulation (NVE simulation): DP-MD, represented by dashed lines. Details on how these results were obtained are provided in the SM.}  
    \label{fig: freq}
\end{figure}

The distance data (Table \ref{tab: nu}) confirms the findings by Ceriotti and coworkers\cite{ceriotti2013nuclear} that density changes primarily alter the average O-O distance. Comparing XC functionals, vdW-cx has the shortest ${d_{O\cdots O'}}$ and the highest density in both simulation types, while PBE presents the opposite behavior. 
Overall, these results suggest that the strength of the hydrogen bond network brings the oxygen atoms closer together, and the density increases despite the small increase in the covalent bond and the larger OHO angles. This is independent of the choice of exchange-correlation functional. Most importantly, it leads to the conclusion that all functionals tend to overbind the water molecules in \iceIH. Considering the quantum nature of the hydrogen atom only makes the overbinding more pronounced.

Finally, we show in Figure \ref{fig: freq} the vibrational spectra of {\iceIH}. The inclusion of NQEs results in a decrease in the bending frequency across all investigated models. However, the most prominent change appears in the stretching frequency range (2500-3500 $\rm{cm^{-1}}$). This shift can be attributed to the fact that these are the modes that are most affected by anharmonic effects. Despite only a slight change in the most probable value for the bond length not changing, the inclusion of NQEs leads to higher populations at larger distances compared to shorter ones. This, in turn, leads to a softening of the modes and a consequent red shift. The shift is in range 218-250 $\rm{cm^{-1}}$, being optB88 the functional with maximum shift (in sequence: optB88, vdW-cx, PBE and SCAN). We note that functionals with van de Waals interactions present the largest shifts, implying stronger H-bonds. NQEs also leads a broadening around 380 $\rm{cm^{-1}}$ in the stretching frequency range in comparison with the DP-MD results for all functionals. Once again, the choice of functional determines the range of stretching frequency, since vdW-cx functional presents the widest range for both DP-PIMD and DP-MD simulations and optB88 has the smallest range for both calculations.

\section{Conclusion}

In summary, we have carried out molecular dynamics simulations of ice structures using a deep potential trained on different exchange-correlation functionals. The inclusion of nuclear quantum effects leads to an overall improvement in the description of the structural properties of ice compared to experiments. In fact, the DP-PIMD radial distribution functions were, in general, more accurate than DP-MD ones due to an improved description of the O-H distribution. The O-O distribution did not present significant changes with the inclusion of NQEs for large distances, however, for distances below 4 {\AA}, the position of the peak shifts slightly for shorter distances. 

In fact, we notice that the density for all exchange and correlation functionals increased as nuclear quantum effects were taken into consideration. By inspecting the different bonds within the system, we identified that the intramolecular O-H bond is slightly weakened, whereas the hydrogen bonds become stronger, and thus the O-O distances decrease, leading to larger densities, further away from experimental values. This leads us to conclude that GGA-type functionals and SCAN tend to overbind the water molecules in ice, leading to a strong H-bond network that is only enhanced by the inclusion of van der Waals interactions and nuclear quantum effects.

\section*{Acknowledgement}
The authors acknowledge financial support from grant numbers \# 2017/10292-0, 2020/09011-9, 	21/14335-0, 23/09820-2, 2023/03658-9, and 23/11751-9
 from the São Paulo Research Foundation (FAPESP), and from CAPES Finance Code 001.  This work used the computational resources from the Centro Nacional de Processamento de Alto Desempenho em São Paulo (CENAPAD-SP), GRID-UNESP and the Santos Dummont Supercomputer (LNCC).

\section{Author Declarations} 

\subsection{Conflict of Interest}

The authors have no conflicts to disclose.

\subsection{Author Contributions} 

{\bf Lucas T. S. de Miranda}: Data curation (equal); Formal analysis (equal); Investigation (equal); Writing – original draft (equal); Writing – review \& editing (equal).  {\bf Márcio S. Gomes-Filho}: Data curation (equal); Formal analysis (equal); Investigation (equal); Writing – original draft (equal); Writing – review \& editing (equal). {\bf Mariana Rossi}: Conceptualization (equal); Formal analysis (equal); Methodology (equal); Writing – original draft (equal); Writing – review \& editing (equal). {\bf Luana S. Pedroza}: Formal analysis (equal); Methodology (equal); Writing – original draft (equal); Writing – review \& editing (equal). {\bf Alexandre R. Rocha}: Conceptualization (equal); Data curation (equal); Formal analysis (equal); Funding acquisition (equal); Methodology (equal); Project administration (equal); Supervision (equal); Writing – original draft (equal); Writing – review \& editing (equal).

\subsection{Data Availability}

The data that support the findings of this study are available from the corresponding author upon reasonable request.


\bibliography{references}

\end{document}


\title{Supplementary Material: Hexagonal ice density dependence on inter atomic distance changes due to nuclear quantum effects}

\author{Lucas T. S. de Miranda}
\affiliation{Institute of Theoretical Physics, São Paulo State University (UNESP), Campus S\~ap Paulo, Brazil}

\author{M\'arcio S. Gomes-Filho}
\affiliation{Centro de Ciências Naturais e Humanas, Universidade Federal do ABC, 09210-580, Santo André, São Paulo, Brazil}

\author{Mariana Rossi}
\affiliation{Max Planck Institute for the Structure and Dynamics of Matter, Luruper Chaussee 149, 22761 Hamburg, Germany}

\author{Luana S. Pedroza}
\affiliation{Instituto de F\'isica, Universidade de S\~ao Paulo, SP 05508-090, Brazil}
\email{luana@if.usp.br}

\author{Alexandre R. Rocha}
\affiliation{Institute of Theoretical Physics, São Paulo State University (UNESP), Campus S\~ao Paulo, S\~ao
 Paulo, Brazil}
\affiliation{Max Planck Institute for the Structure and Dynamics of Matter, Luruper Chaussee 149, 22761 Hamburg, Germany}

\email{alexandre.reily@unesp.br}

\maketitle


\section{Computational details}
 
\begin{figure}[!htb]
    \centering
    \includegraphics[width=\textwidth]{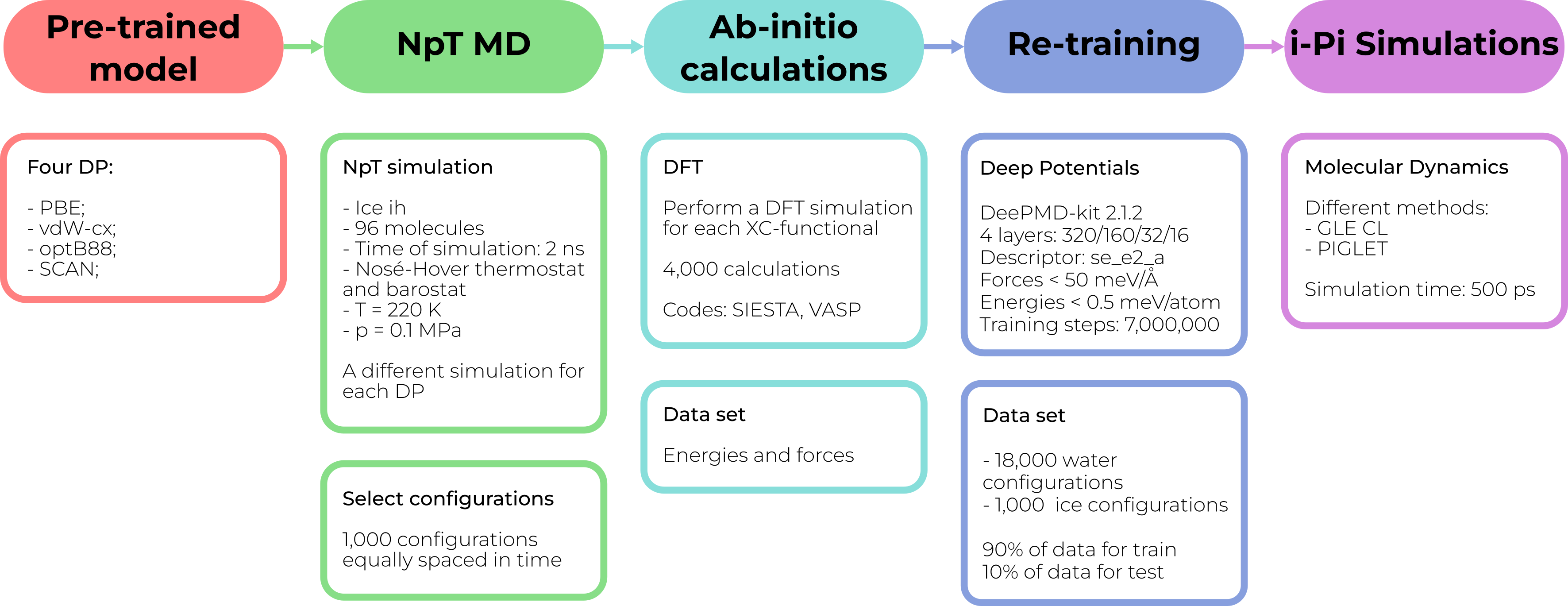}
    \caption{Workflow of the development and use of a deep potential model to investigate the properties of hexagonal ice.}
    \label{fig:worflow-SI}
\end{figure}

Our simulation protocol for developing an accurate deep neural network potential (Deep Potential - DP) to investigate the properties of hexagonal ice is outlined in Figure~\ref{fig:worflow-SI}. 
Essentially, the protocol involves five key stages. The first stage encompassed pre-training models for PBE\cite{perdew1996generalized}, vdW-cx\cite{BH}, SCAN\cite{SCAN-original}, and optB88-vdW\cite{optB88} XC-functionals based on the DFT dataset by Torres~\textit{et. al.}\cite{Torres21, GomesFilho23}. 

The second stage involved conducting isobaric-isothermal ensemble (NPT) molecular dynamics simulations using the pre-trained models to generate ice configurations for inclusion in the training dataset.
In this step, we performed a long simulation, lasting up to 2 ns, using a time step of 0.25 fs, at a fixed temperature of $T = 220$ K and pressure of $P = 0.1$ MPa.
The Nose-Hoover\cite{tuckerman1993efficient} thermostat and barostat were employed with coupling parameters as suggested by LAMMPS.\cite{Lammps22} 
We then selected $10^3$ configurations, uniformly spaced in time, for inclusion in the training dataset, meaning  one configuration was selected  for every 2 ps of overall dynamics.

In the third stage, DFT energies and forces were obtained for each ice configuration and each XC-functional. For PBE, vdW-cx and optB88-vdW single-point calculations for each ice configuration, the SIESTA\cite{artacho2018siesta} code was used. We employed norm-conserved pseudopotentials\cite{troullier1991efficient} and a 500 Ry mesh cutoff. An optimized double-$\zeta$ basis\cite{corsetti2013optimal} set was used for the valence electrons. The single-point simulations for SCAN were carried out with VASP\cite{kresse1996efficient}, using a plane wave basis set with an energy cutoff of 118 Ry, where the core-valence interaction was treated by the projected augmented wave  method.\cite{kresse1999ultrasoft}

The fourth stage focused on retraining each DP model using DeePMD-kit\cite{wang2018deepmd, DeepMDreview2023:JCP, lu202186} incorporating the new ice configurations to the training data set. 
The Deep Potential-Smooth Edition descriptor was used, where all  relative coordinates (distances and angles) were used to build the descriptor\cite{zhang2018end}, within a radius cutoff of 6.0 {\AA} and a smoothing ratio of 0.5 {\AA}. The possible maximum number of neighbors in the cut-off radius is set to be 46 for oxygen  and 92 for hydrogen atoms.
The fitting network architecture consists of four layers with 320, 160, 32, and 16 neurons in each layer, respectively, and the hyperbolic tangent activation function was selected.
The loss function is composed by the mean squared errors of the energies and forces, and was minimized with the Adam stochastic gradient descent method for $7\times10^6$ training steps. Loss function parameters are present in Table \ref{tab: loss-SI}. 

\begin{table}[!htb]
    \centering
        \begin{tabular*}{.5\columnwidth}{@{\extracolsep{\fill}}cc@{}}\hline\hline
   Loss parameter & Value \\ \hline\hline
   start\_pref\_e & 0.02\\
   limit\_pref\_e & 8.00 \\
   start\_pref\_f & 1500\\
   limit\_pref\_f & 1.00\\\hline\hline
    \end{tabular*}
    \caption{Loss function parameters used for training.}
    \label{tab: loss-SI}
\end{table}

We selected 90\% of the configurations as the training set and 10\% as the test set, corresponding to 17100 (training) and 1900 (testing) structures. For training these NN-based potentials, we used a single AMD Epyc 7662 with a NVIDIA Tesla A100. Each potential took 22 hours to complete the training steps. After completing the training process, the model was compressed to improve the computational performance of our DP-MD simulations.\cite{Lu2022:dpcompress}

The fifth and final stage involved the use of the trained DP models to conduct Deep Potential Molecular Dynamics (DP-MD) and Deep Potential Path Integral Molecular Dynamics (DP-PIMD).
We simulated a proton-disordered structure of \iceIH~ consisting of 96 water molecules.  For all simulations, the systems were first equilibrated over 100 ps, followed by an additional 500 ps of production simulations, with a time step of 0.25 fs, and the data were collected every 0.05 ps. 
We chose to investigate a specific point on the water phase diagram (T = 220 K and P = 0.1 MPa). All simulations were performed using the i-PI code.\cite{kapil2019pi} We use the Generalized Langevin Equation (GLE) thermostat\cite{ceriotti2010colored} for DP-MD (optimal sampling) and for DP-PIMD simulations we used the PIGLET\cite{ceriotti2012efficient, ceriotti2009nuclear} method with 8 beads and the ratio $\rm{\omega_{max}/\omega_{min}=10^4}$ to include the nuclear quantum effects (NQEs). We investigated the vibrational properties of ice Ih using two types of simulations: classical molecular dynamics in the NVE ensemble and a path-integral approach with thermostatted ring-polymer molecular dynamics (TRPMD).

\section{Validation of deep potentials}

To assess the quality of the DP models, we conducted two validations procedures: \textit{a priori}, where we measured the error in energy and force between the predictions made by the DP models and the DFT reference, using the testing dataset.

\begin{figure}[htbp]
    \centering
    \includegraphics[width=\textwidth]{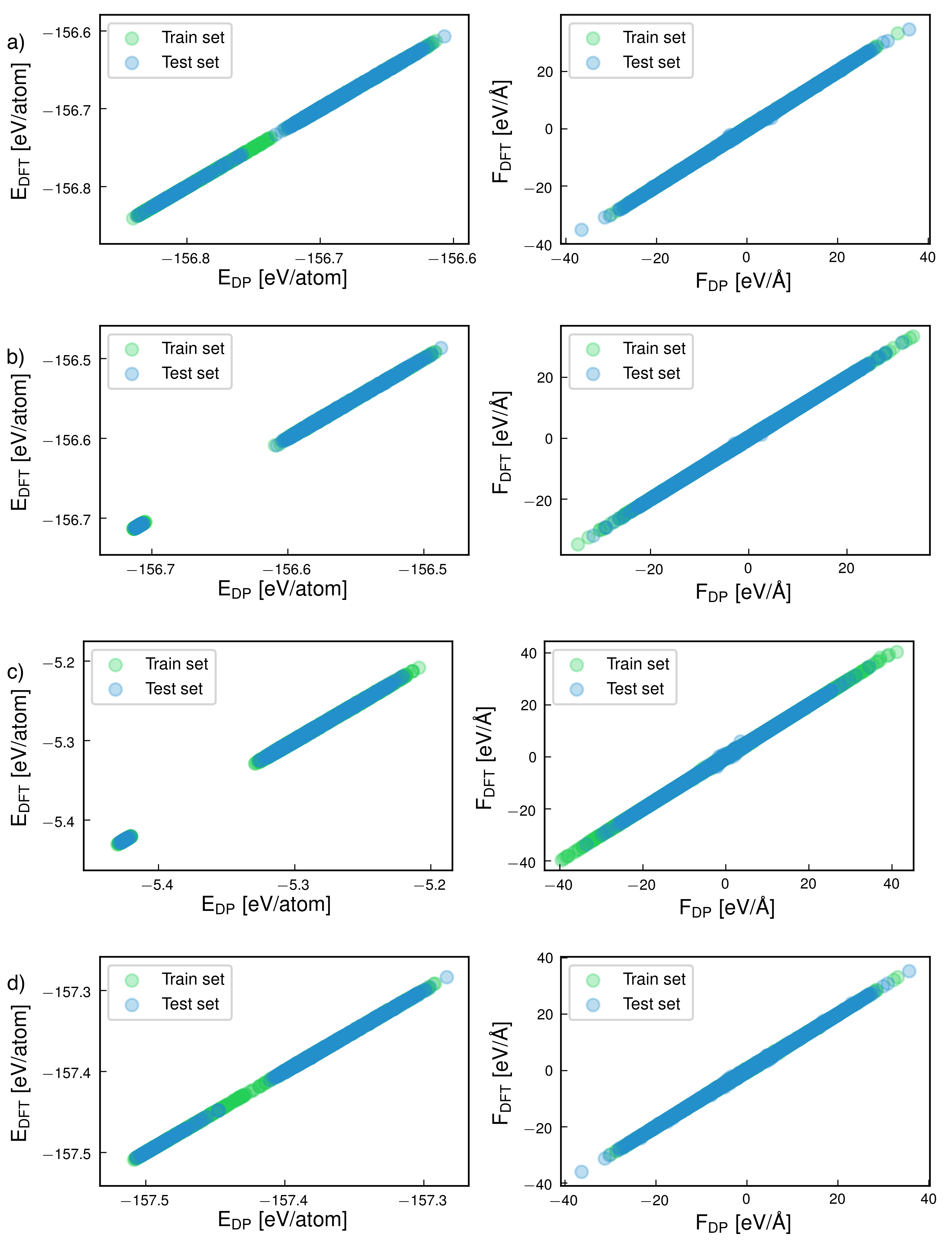}
    \caption{Parity plots for energies per atom and forces for each DP model: a) PBE, b) vdW-cx, c) SCAN and d) optB88-vdW.}
    \label{fig: rsme-SI}
\end{figure}

In Figure \ref{fig: rsme-SI}, we present parity plots for energies (left panel) and forces (right panel) for each DP model (from top to bottom: PBE, vdW-cx, SCAN, and optB88-vdW) for both the training and testing data sets. As observed, all models demonstrate a good fit to the datasets. Table \ref{tab: RMSE} provides the root mean square error (RMSE) and mean absolute error (MAE). The obtained RMSEs for the test sets are approximately 45 meV/{\AA} and 0.5 meV/atom for forces and energies, respectively.

The second validation procedure,  \textit{a posteriori}, examines whether the coordinates generated from DP-MD simulations produce structures with reasonable errors in energy and force when compared to DFT calculations. 

\begin{figure}[htbp]
    \centering
    \includegraphics[width=\textwidth]{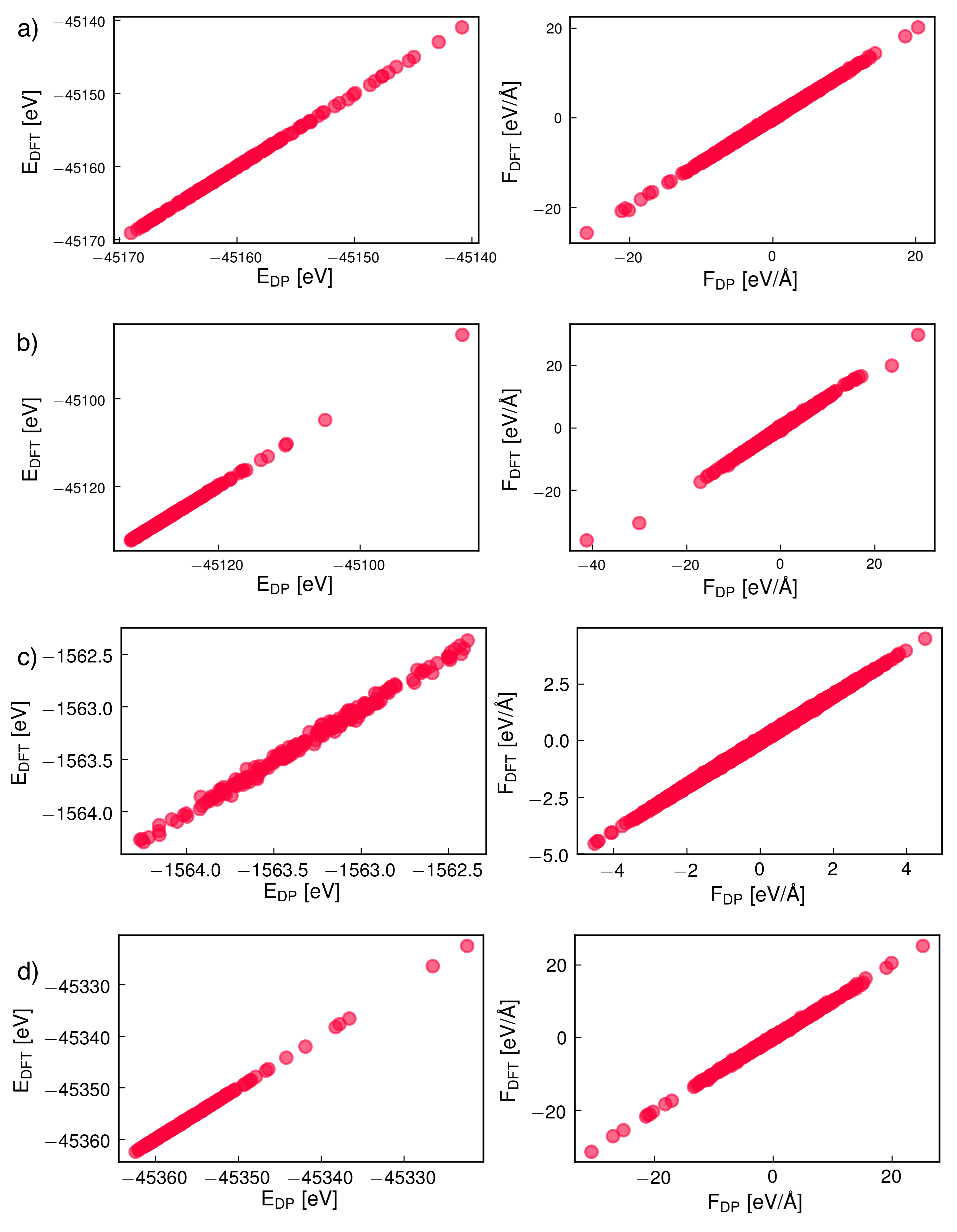}
    \caption{Parity plots for energies and forces for each DP model for configurations collected on production run: a) PBE, b) vdW-cx, c) SCAN and d) optB88-vdW. }
    \label{fig: rsme-post-SI}
\end{figure}

In this approach, the DP models were tested on 200 {\iceIH} configurations with different densities obtained from the production run.
Subsequently, we computed the DFT energies and forces for these configurations. The resulting errors between DP and DFT results are depicted in \ref{fig: rsme-post-SI} and detailed in Table~\ref{tab: RMSE-post}. Notably, we achieved lower error values in this \textit{a posteriori} validation, indicating the accuracy of the fitted potentials. Given these reduced error values, there was no need to incorporate additional training data points in the training datasets.

\begin{table}[!h]
    \centering
    \begin{tabular*}{.95\textwidth}{@{\extracolsep{\fill}}ccccc@{}}\hline\hline
    & \multicolumn{2}{c}{Force (meV/\AA)} & \multicolumn{2}{c}{Energy (meV/atom)}\\ \hline\hline
     DP model &  RMSE & MAE & RMSE & MAE \\\hline\hline
         PBE & 43.7 & 33.1 & 0.49 & 0.39\\
         vdW-cx & 44.6 & 33.5 & 0.52 & 0.40\\
         SCAN & 46.9 & 35.2 & 0.56 & 0.44\\
         optB88-vdW & 44.9 & 33.9 & 0.50 & 0.40\\\hline\hline
    \end{tabular*}
    \caption{Values of the RMSE and MAE on test dataset.}
    \label{tab: RMSE}
\end{table}

\begin{table}[!h]
    \centering
    \begin{tabular*}{.95\textwidth}{@{\extracolsep{\fill}}ccccc@{}}\hline\hline
    & \multicolumn{2}{c}{Force (meV/\AA)} & \multicolumn{2}{c}{Energy (meV/atom)}\\ \hline\hline
     DP model &  RMSE & MAE & RMSE & MAE \\\hline\hline
         PBE & 38.9 & 28.1 & 0.25 & 0.21\\
         vdW-cx & 46.6 & 31.1 & 0.34 & 0.29\\
         SCAN & 23.7 & 18.3 & 0.13 & 0.10\\
         optB88-vdW & 39.6 & 27.6 & 0.20 & 0.15\\\hline\hline
    \end{tabular*}
    \caption{Values of the RMSE and MAE for \textit{a posteriori} configurations collected during the production run.}
    \label{tab: RMSE-post}
\end{table}

\section{The choice of PIGLET method}

All simulations were performed using the i-PI code.\cite{kapil2019pi} We used the Generalized Langevin Equation (GLE) thermostat\cite{ceriotti2010colored} for DP-MD (optimal sampling) calculations, and for the DP-PIMD simulations we used the PIGLET\cite{ceriotti2012efficient, ceriotti2009nuclear} method with 8 beads and the ratio $\rm{\omega_{max}/\omega_{min}=10^4}$ to include the NQEs. 

\begin{figure}[htbp]
     
     \includegraphics{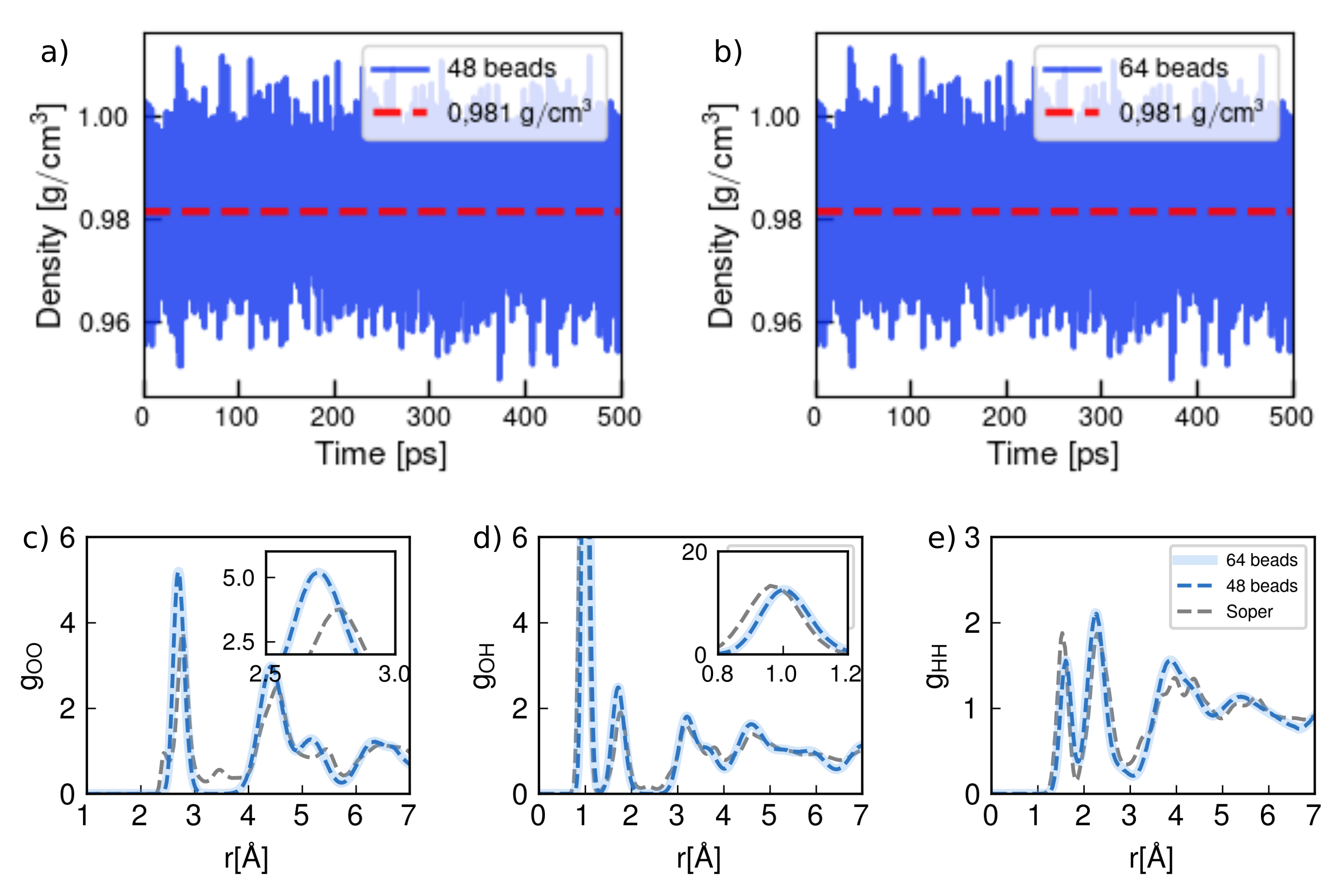} \\
    \caption{Time evolution of the ice density obtained from PILE-L simulations using a) 48 beads and b) 64 beads. RDFs for c oxygen-oxygen (g$\rm_{OO}$), d) oxygen-hydrogen (g$\rm_{OH}$) and e) hydrogen-hydrogen (g$\rm_{HH}$) (bottom panel) for different numbers of beads are presented for PILE-L results for 48 and 64 beads. The experimental results are also shown~\cite{soper2000radial}.}
    \label{fig:pilel d}
\end{figure}

It is important to mention that we tested the convergence of the results with respect to the number of beads by conducting simulations using 48 and 64 beads   with the PILE-L method\cite{ceriotti2010efficient} for the optB88-vdW deep potential. As shown in Fig. \ref{fig:pilel d} we obtained similar results for the ice density and the radial distribution functions (RDFs)  with both bead counts. This indicates that convergence is achieved with 48 beads.

\begin{figure}[htbp]
     \includegraphics{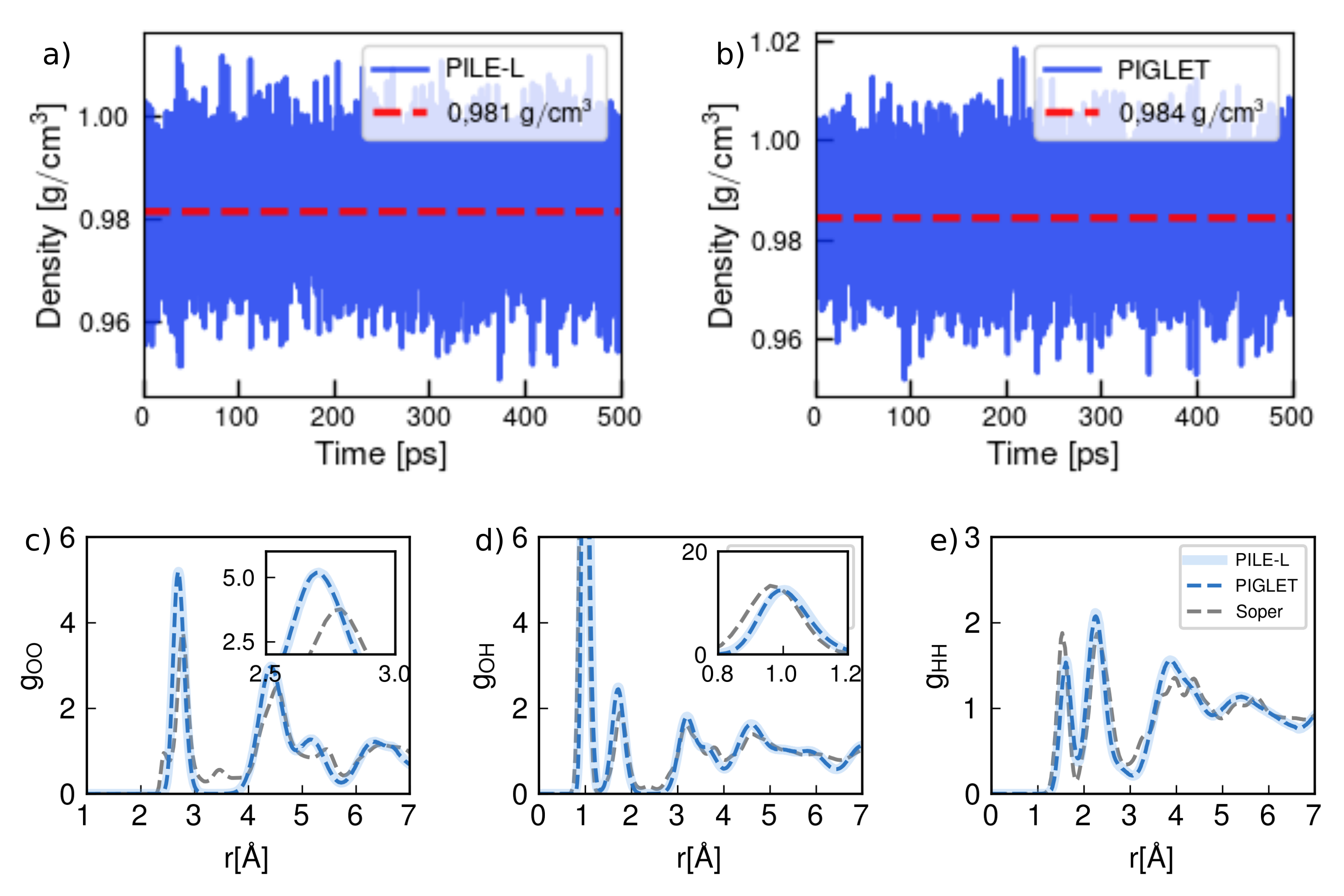}
    \caption{Time evolution of the ice density obtained from PILE-L simulations using a) PILE-L method with 48 beads and b) PIGLET method with 8 beads. RDFs for c) oxygen-oxygen (g$\rm_{OO}$), d) oxygen-hydrogen (g$\rm_{OH}$) and e) hydrogen-hydrogen (g$\rm_{HH}$) (bottom panel) are presented for PILE-L and PIGLET results. The experimental results are also shown~\cite{soper2000radial}.}
    \label{fig:pilel_vs_piglet d} 
\end{figure}

In this context, we compare the simulation results obtained using both the PILE-L method with 48 beads and the PIGLET method with 8 beads, as illustrated in Fig. \ref{fig:pilel_vs_piglet d}. The PIGLET method with 8 beads provides results similar to those of the PILE-L method. Given that the PIGLET method offers a lower computational cost compared to PILE-L, we have chosen to conduct all DP-PIMD simulations using the PIGLET method.

\section{Vibrational Properties}

To investigate the vibrational properties of \iceIH, we conducted two types of simulations using the i-PI code\cite{kapil2019pi}. The first was a classical molecular dynamics (MD) simulation in which the number of particles (N), volume (V), and energy (E) were held constant (NVE ensemble). The second simulation employed a path-integral-based approach, specifically thermostatted ring-polymer molecular dynamics (TRPMD)\cite{Rossi14:JCP,Rossi14:JCPcommun}.

\begin{figure}[htbp]
    \centering
    \includegraphics{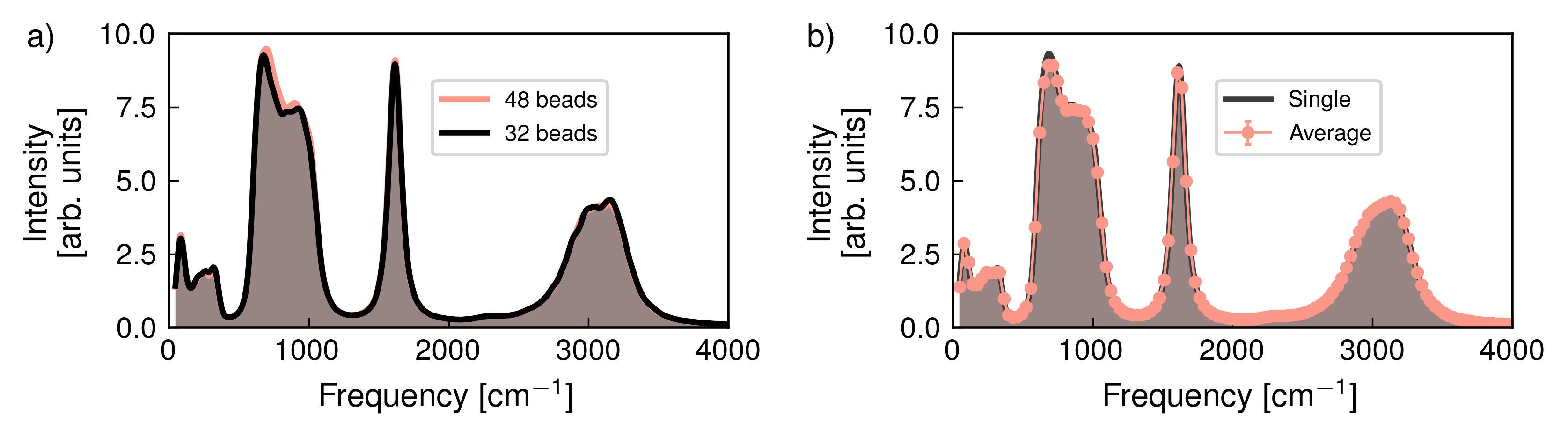}
    \caption{Comparison of vibrational spectra obtained from TRPMD simulations using a) 32 and 48 beads and b) from a single 100 ps simulation and the average of 10 independent 10 ps simulations for PBE.}
    \label{fig:beads-SI}
\end{figure}

Figure~\ref{fig:beads-SI} a) shows the vibrational spectra of \iceIH~at 220 K, obtained via Fourier transform of the velocity autocorrelation function from TRPMD simulations using 32 and 48 beads.
The parameter $\lambda$ was set to $1/2$, as recommended by Rossi and collaborators\cite{Rossi14:JCP}. Each trajectory ran for 10 ps with a time step of 0.25 fs, starting from the final configuration of the DP-PIMD production run (500 ps). As shown in the figure, the simulation using 32 beads has converged. Therefore,  the subsequent TRPMD simulations were performed using 32 ring polymer beads. 

The final spectral results, presented in the main text, were obtained by averaging 10 independent 10 ps trajectories, each initiated from an equilibrated configuration chosen from the 500 ps production run of the DP-PIMD simulations. In Fig.~\ref{fig:beads-SI} b), we present the spectrum obtained from a single 100 ps trajectory alongside the averaged spectrum. As can be observed, averaging multiple short simulation runs yields comparable results to those from a single long trajectory.

The vibrational spectra obtained from the classical NVE simulations were performed in exactly the same way;  by averaging 10 independent 10 ps trajectories with a time step of 0.25 fs, but the equilibrated initial configurations were chosen from the 500 ps production run of the DP-MD simulations.

\section{Complementary results}

\begin{figure}[htbp]
    \centering
    \includegraphics[scale=0.9]{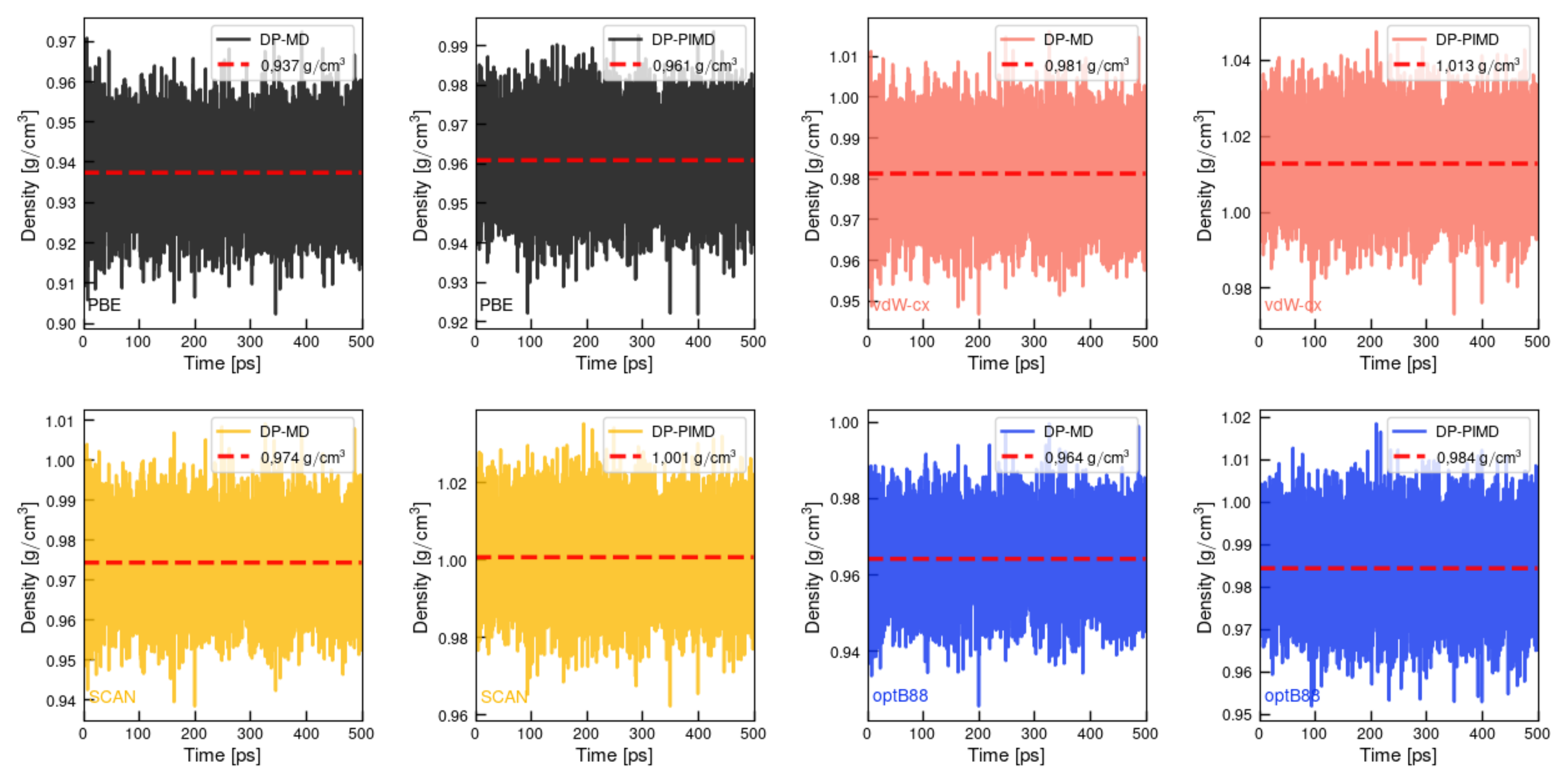}
    \caption{Results of DP-MD and DP-PIMD simulations showing the time evolution of the density for each DP model. Note that the reference for each DP potential is displayed in the corresponding panel.}
    \label{fig:density-SI}
\end{figure}

\begin{figure}[htbp]

\includegraphics{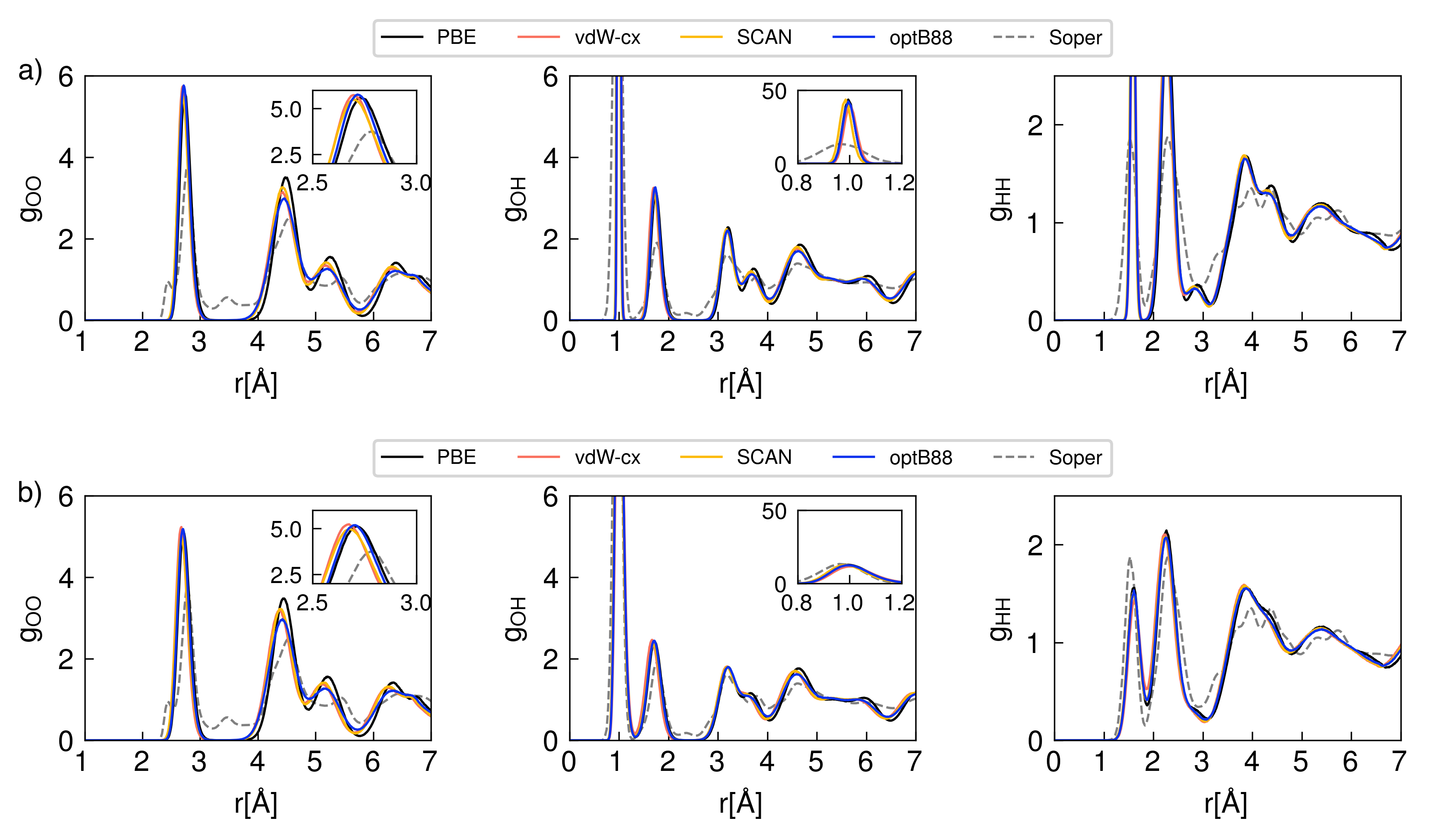}
\caption{RDFs for oxygen-oxygen (g$\rm_{OO}$), oxygen-hydrogen (g$\rm_{OH}$), and hydrogen-hydrogen (g$\rm_{HH}$) for a) DP-MD and b) DP-PIMD simulations for each XC-functional. The experimental results are also shown~\cite{soper2000radial}.}
\label{fig:rdfs}
\end{figure}

As detailed in the main text, we conducted an NPT simulation at T = 220 K and p = 0.1 MPa to observe the density of {\iceIH} for each potential. 
The equilibrium density was obtained as the average density over 500 ps of production. Figure \ref{fig:density-SI} presents the time evolution of density for each model in both DP-MD and DP-PIMD simulations. The horizontal dashed lines indicate the equilibrium densities. Additionally, we provide a comparison between the RDFs obtained from DP-MD and DP-PIMD for each XC-functional in Fig. \ref{fig:rdfs}.

\begin{figure}[htbp]
\includegraphics{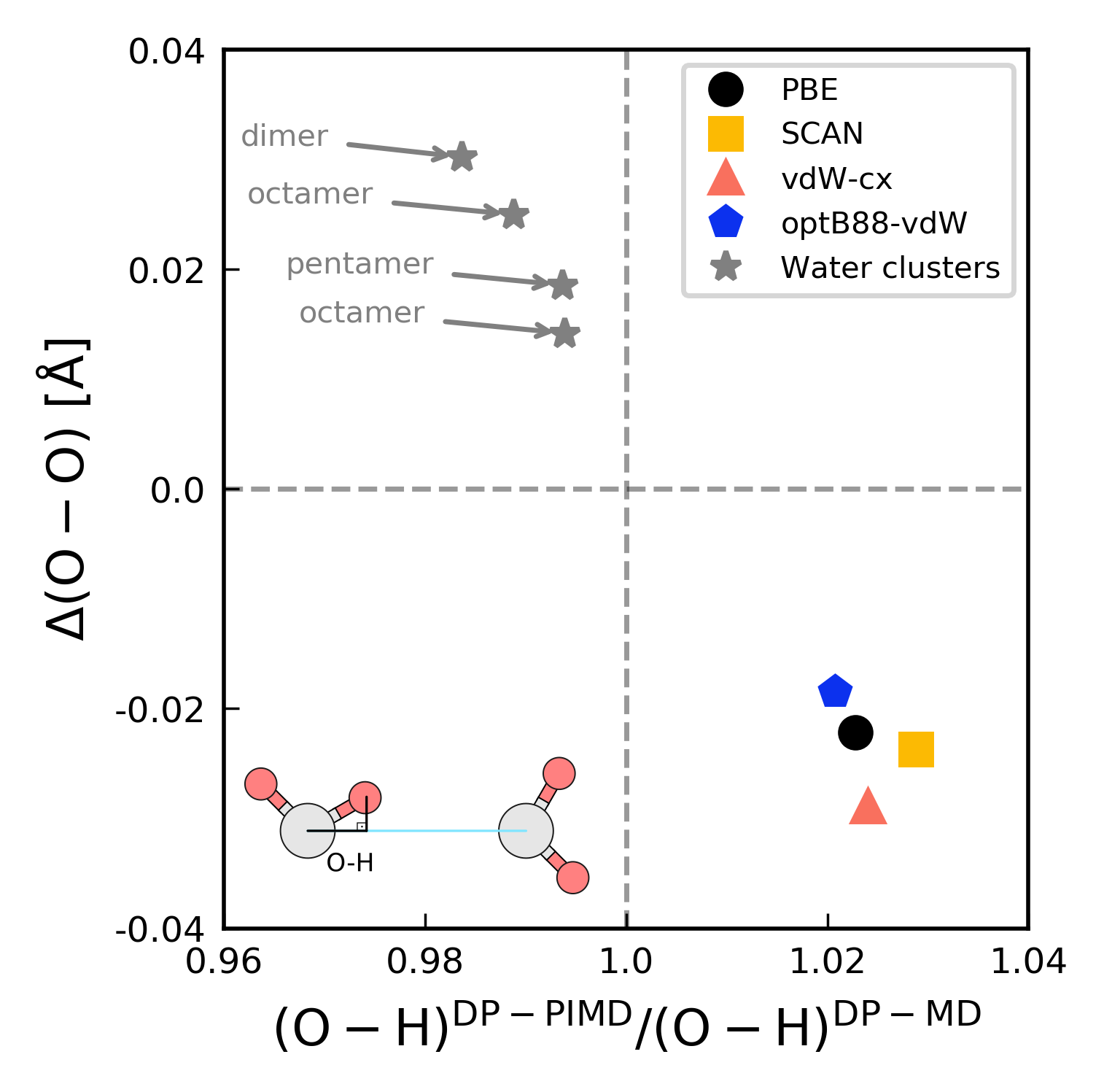}
\caption{Relationship between the projection of O-H bonds in the O-O direction and the O-O distance. The values of water clusters were theoretically obtained by Li.\cite{doi:10.1073/pnas.1016653108}}
\label{fig:proj}
\end{figure}

Li \textit{et. al.}\cite{doi:10.1073/pnas.1016653108} observe from a theoretical point of view the relationship between the projection of O-H bonds in the O-O direction and the O-O distance, as shown by the inset of Figure \ref{fig:proj}, when the NEQs were included. In their analysis, with the increase of the projection of O-H bond, the O-O distances decrease. The opposite behavior was reported when the projection enhanced. Since NQEs tend to weaken the weak bonds and strengthen the strong bonds, all functionals strengthen the H-bonds and reproduce the observations about the decrease of O-O distances as seen in Figure \ref{fig:proj}.

\pagebreak

\bibliography{references2}